\RequirePackage{fixltx2e}
\documentclass[twoside,english,journal, onecolumn, draftclsnofoot, 10pt]{IEEEtran}
\usepackage[T1]{fontenc}
\usepackage[latin9]{inputenc}
\usepackage{babel}
\usepackage{float}
\usepackage{graphicx}
\usepackage[unicode=true,pdfusetitle,
 bookmarks=true,bookmarksnumbered=false,bookmarksopen=false,
 breaklinks=false,pdfborder={0 0 1},backref=false,colorlinks=false]
 {hyperref}
\usepackage{breakurl}

\makeatletter

\newcommand{\lyxdot}{.}

\setkeys{Gin}{width=\linewidth}
\usepackage{cite}

\@ifundefined{showcaptionsetup}{}{%
 \PassOptionsToPackage{caption=false}{subfig}}
\usepackage{subfig}
\makeatother

\begin{document}

\title{Waveform Design for 5G and Beyond}

\author{\IEEEauthorblockN{Ali Fatih Demir\textsuperscript{1}, Mohamed Elkourdi\textsuperscript{1}, Mostafa Ibrahim\textsuperscript{2}, Huseyin Arslan\textsuperscript{1,2}}\\
\IEEEauthorblockA{\textsuperscript{1}University of South Florida, Department of Electrical Engineering, Tampa, FL, USA}\\
\IEEEauthorblockA{\textsuperscript{2}Istanbul Medipol University, Department of Electrical and Electronics Engineering, Istanbul, Turkey\\}
\thanks{\textbf{This is the pre-peer reviewed version of the article which has been published in final form at [https://onlinelibrary.wiley.com/doi/abs/10.1002/9781119333142.ch2].
This article may be used for non-commercial purposes in accordance with Wiley Terms and Conditions for Use of Self-Archived Versions.}}}
\maketitle

\section{Introduction \label{sec: I}}

The standardization activities of wireless mobile telecommunications
have begun with analog standards that were introduced in the 1980s,
and a new generation develops almost every 10 years to meet the exponentially
growing market demand. The leap from analog to digital started in
second-generation (2G) systems, along with the use of mobile data
services. The 3G digital evolution enabled video calls and global
positioning system (GPS) services on mobile devices. The 4G systems
pushed the limits of data services further by better exploiting the
time\textendash frequency resources using orthogonal frequency-division
multiple access (OFDMA) as an air interface \cite{elkourdi2016}.
Recently, the International Telecommunications Union (ITU) has defined
the expectations for 5G \textbf{\cite{ITU-R}}, and the study of the
next-generation wireless system is in progress with a harmony between
academia, industry, and standardization entities to accomplish its
first deployment in 2020. 

5G is envisioned to improve major key performance indicators (KPIs),
such as peak data rate, spectral efficiency, power consumption, complexity,
connection density, latency, and mobility, significantly. Furthermore,
the new standard should support a diverse range of services all under
the same network \textbf{\cite{3gppQualReq}}. The IMT-2020 vision
defines the use cases into three main categories as enhanced mobile
broadband (eMBB), massive machine-type communications (mMTC), and
ultra-reliable low-latency communications (URLLC) featuring 20 Gb/s
peak data rate, 10\textsuperscript{6}/km\textsuperscript{2} device
density, and less than 1 ms latency, respectively \cite{zhang2016}.
A flexible air interface is required to meet these different requirements.
As a result, the waveform, which is the main component of any air
interface, has to be designed precisely to facilitate such flexibility
\cite{3gppHuaweiReq}. 

This chapter aims to provide a complete picture of the ongoing 5G
waveform discussions and overviews the major candidates. The chapter
is organized as follows: Section \ref{sec:II} provides a brief description
of the waveform and reveals the 5G use cases and waveform design requirements.
Also, this section presents the main features of CP-OFDM that is currently
deployed in 4G LTE systems. CP-OFDM is the baseline of the 5G waveform
discussions since the performance of a new waveform is usually compared
with it. Section \ref{sec:III} examines the essential characteristics
of the major waveform candidates along with the related advantages
and disadvantages. Section \ref{sec:IV} summarizes and compares the
key features of the waveforms. Finally, Section \ref{sec:V} concludes
the chapter. 

\section{Fundamentals of the 5G Waveform Design \label{sec:II}}

\subsection{Waveform Definition }

\begin{figure}[t]
\centering\includegraphics[width=0.7\columnwidth]{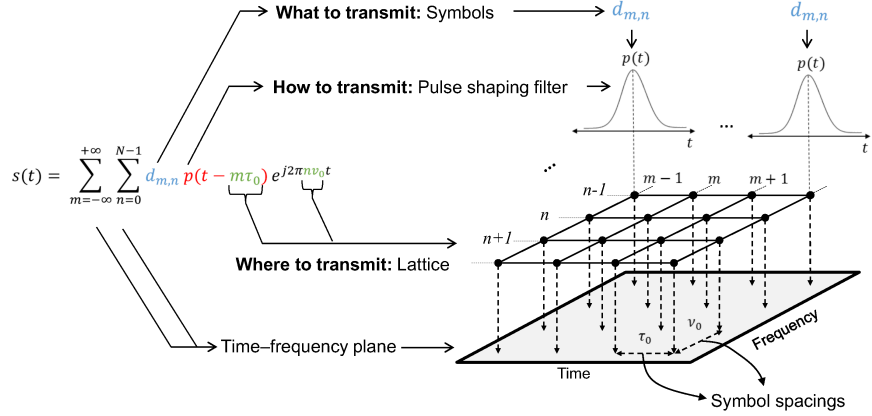}
\centering{}\caption{Waveform definition \cite{sahin2014}. \label{fig:Waveform-Definition.}}
\end{figure}

The waveform defines the physical shape of the signal that carries
the modulated information through a channel. The information is mapped
from the message space to the signal space at the transmitter, and
a reverse operation is performed at the receiver to recover the message
in a communications system. The waveform, which defines the structure
and shape of the information in the signal space, can be described
by its fundamental elements: symbol, pulse shape, and lattice, as
shown in Fig \ref{fig:Waveform-Definition.}. The symbols constitute
the random part of a waveform whereas the pulse shape and the lattice
form the deterministic part. 
\begin{itemize}
\item \emph{\textit{Symbol:}} A symbol is a set of complex numbers in the
message space that is generated by grouping a number of bits together.
The number of bits grouped within one symbol determines the modulation
order that has a high impact on the throughput. 
\item \emph{\textit{Pulse Shape:}} The form of the symbols in the signal
plane is defined by the pulse shaping filters. The shape of the filters
determines how the energy is spread over the time and frequency domains
and has an important effect on the signal characteristics. 
\item \emph{\textit{Lattice:}} The lattice is generated by sampling the
time\textendash frequency plane, and the locations of samples define
the coordinates of the filters in the time\textendash frequency grid.
The lattice geometry might present different shapes such as rectangular
and hexagonal according to the formation and distances between the
samples. Furthermore, the lattice can be exploited by including additional
dimensions such as space domain.
\end{itemize}

\subsection{5G Use Cases and Waveform Design Requirements}

The new radio for 5G should support a wide range of services as discussed
in Section \ref{sec: I}. Primarily, the applications that require
larger bandwidth and spectral efficiency falls into eMBB category,
whereas the ones that have a tight requirement for device battery
life falls into mMTC. Usually, the industrial smart sensors \cite{akyildiz2002}
or medical implants \cite{demir2016a} operate several years without
the demand for maintenance and hence low device complexity and high
energy efficiency are crucial for these mMTC services. Furthermore,
the mission-critical applications such as remote surgery \cite{demir2016b}
or self-driving vehicles \cite{urmson2008} are represented in URLLC.
The key requirements that are associated with each of these use cases
are summarized in Table \ref{tab:The-5G-UseCases} and the following
design criteria are essential to meet these requirements of 5G: 

\begin{table}[b]
\caption{The 5G use cases \label{tab:The-5G-UseCases}}

\centering\includegraphics[width=0.4\columnwidth]{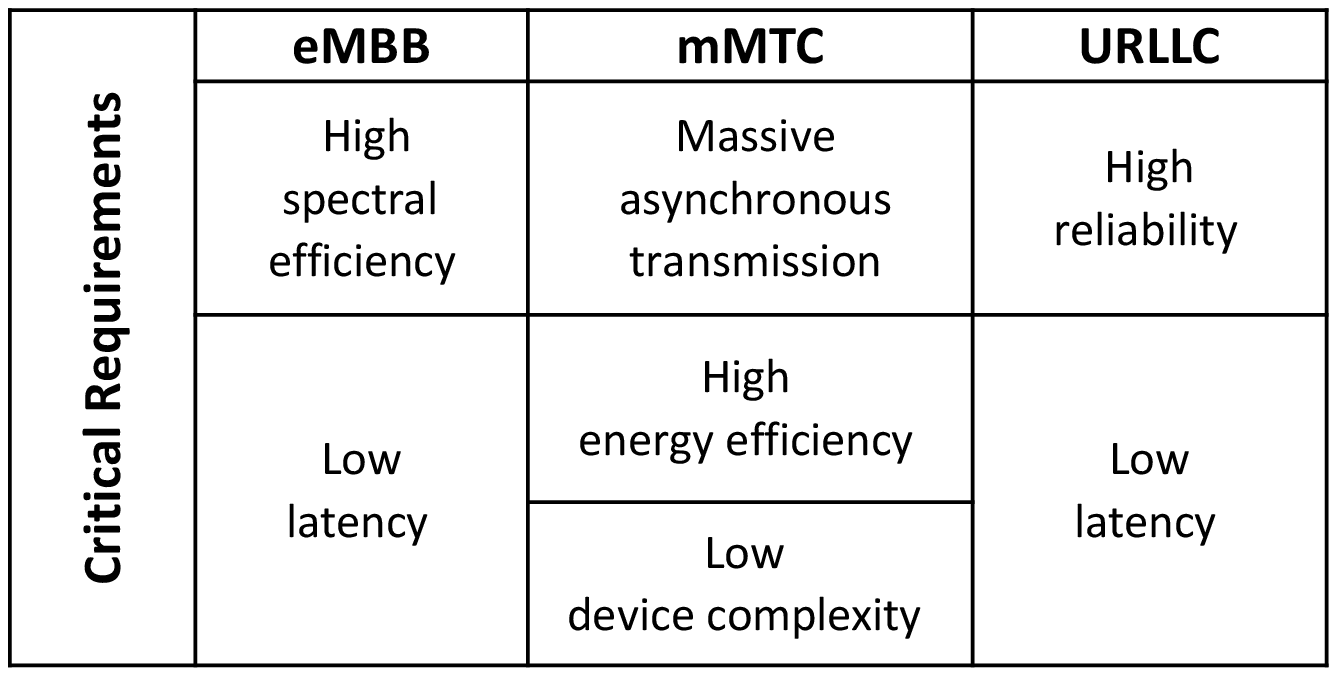}
\end{table}
\begin{itemize}
\item \emph{\textit{High Spectral Efficiency:}} The modulation order, type
of pulse shaping filters, and density of the lattice play an important
role in determining the spectral efficiency. The guard units in time
or frequency domains and other extra overheads decrease the spectral
efficiency, which is critical especially for the eMBB type of communications.
Furthermore, the multi-antenna techniques \cite{hafez2015} such as
beamforming and massive MIMO is another crucial aspect to utilize
the lattice more efficiently. However, the self-ISI and self-ICI of
a waveform prevents to apply MIMO techniques in a straight-forward
way and increases complexity significantly.
\item \emph{\textit{Low Latency:}} 5G targets a latency less than 1 ms
for the URLLC applications. This goal can be managed by shortening
the transmission time interval (TTI) or increasing the sub-carrier
spacing. However, the latter approach increases the relative CP overhead
for a given TTI. Also, the localization in time is critical, and shorter
filter/window durations are needed to fulfill this requirement. 
\item \emph{\textit{High Reliability:}} The reliability is evaluated by
bit error rate (BER) or block error rate (BLER), and it is extremely
important for mission-critical communications where errors are less
tolerable. In addition, the re-transmissions due to errors cause an
increase in latency, and hence high reliable links are desirable to
provide low latency as well. 
\item \emph{\textit{Massive Asynchronous Transmission:}} It is envisioned
that there will be a huge number of nodes communicating over the 5G
network for mMTC services. To maintain the synchronicity, excessive
overhead is required for these applications. However, it decreases
the spectral efficiency significantly. The waveforms that have strict
synchronization requirements to achieve interference-free communications
are not suitable for mMTC applications. Therefore, the waveforms that
are well localized in the multiplexing domain are more suitable to
relax the synchronization requirement for these type of applications.
\item \emph{\textit{Low Device Complexity:}} The computational complexity
is another critical metric of the waveform design and depends on the
number of operations required at the transmitter or receiver. Additional
windowing, filtering, and interference cancellation algorithms increase
complexity substantially, and the system designer should consider
it to design a cost- and energy-efficient transceivers.
\item \emph{\textit{High Energy Efficiency:}} Low computational complexity
and low peak-to-average power ratio (PAPR) provides high energy efficiency.
PAPR is a statistical metric that is evaluated by complementary cumulative
distributive function (CCDF) of the signal. Low PAPR is required to
operate power amplifiers (PAs) efficiently, which are one of the most
energy-hungry components in a transceiver.
\end{itemize}

\subsection{The Baseline for 5G Waveform Discussion: CP-OFDM}

\begin{figure}[t]
\centering\includegraphics[width=0.8\columnwidth]{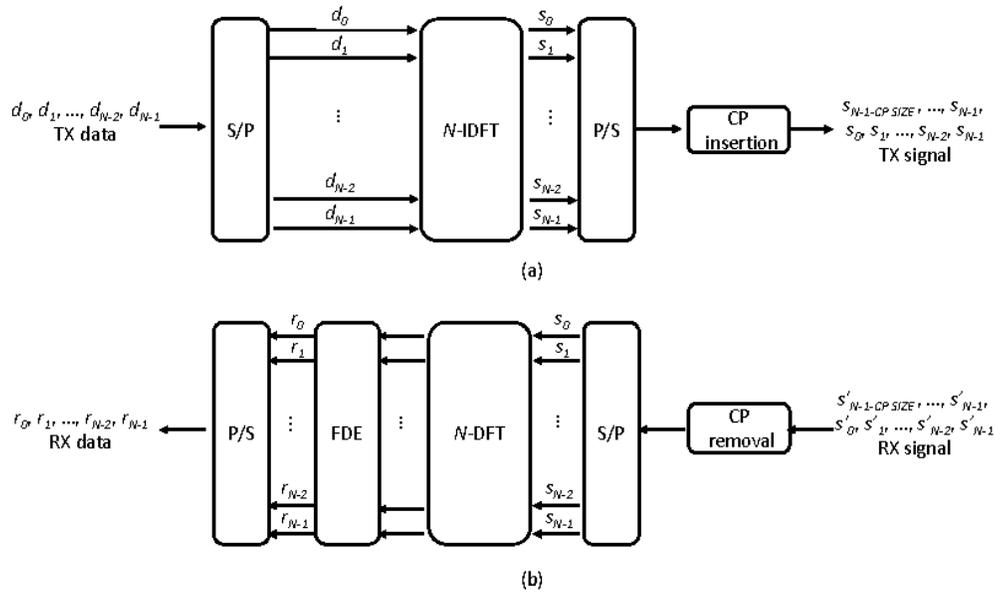}
\centering{}\caption{CP-OFDM block diagram. (a) Transmitter. (b) Receiver.\label{fig:OFDM-Block}}
\end{figure}

Orthogonal frequency-division multiplexing (OFDM) is the most popular
multicarrier modulation scheme that is currently being deployed in
many standards such as the downlink of 4G LTE and the IEEE 802.11
family \cite{hwang2009}. Its primary advantage over the single-carrier
transmission schemes is its ability to cope with frequency selective
channels for broadband communications. The data is divided into parallel
streams, and each is modulated with a set of narrow subcarriers. The
bandwidth of each subcarrier is set to be less than the coherence
bandwidth of the channel. Hence, each subcarrier experiences a single-tap
flat fading channel that can be equalized in the frequency domain
with a simple multiplication operation. Also, OFDM systems utilize
the spectrum in a very efficient manner due to the orthogonally overlapped
subcarriers and allow flexible frequency assigning. A discrete OFDM
signal on baseband is expressed as follows: 

\begin{equation}
s_{OFDM}[k]=\sum_{n=0}^{N-1}d_{n}e^{j2\pi k\frac{n}{N}}\label{eq:Eq1-OFDM}
\end{equation}
where $d_{n}$ is the complex data symbol at subcarrier $n$, and
$N$ represents the total number of subcarriers. OFDM can easily be
implemented by the inverse fast Fourier transform (IFFT) algorithm.
Afterward, the cyclic prefix (CP) is added by copying the last part
of the IFFT sequence and appending it to the beginning as a guard
interval. The CP length is determined based on the maximum excess
delay of the channel to alleviate the effect of inter symbol interference
(ISI). However, it is hard-coded in 4G LTE and does not take into
account the individual user\textquoteright s channel delay spread.
As a result, the fixed guard interval leads to a degradation in the
spectral efficiency. Furthermore, the CP yields to handling the interference
in a multipath environment by ensuring circularity of the channel
and by enabling easy frequency-domain equalization (FDE). OFDM partly
diminishes the inter carrier interference (ICI) as well by setting
the subcarrier spacing according to the maximum Doppler spread. A
block diagram of conventional CP-OFDM transmitter and receiver is
shown in Fig. \ref{fig:OFDM-Block}.

\begin{figure}[b]
\centering\includegraphics[width=0.5\columnwidth]{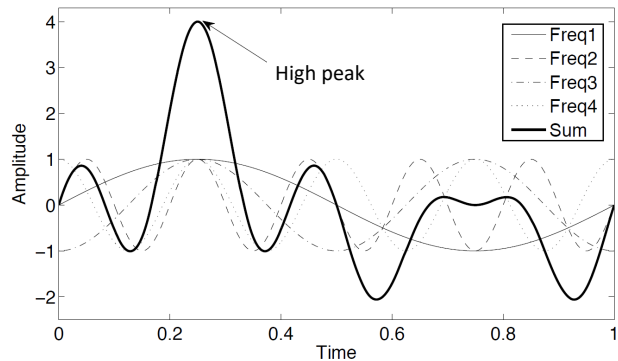}
\centering{}\caption{The PAPR problem in multicarrier schemes \cite{rahmatallah2013}.\label{fig:PAPR}}
\end{figure}

A major disadvantage of any multicarrier system, including CP-OFDM,
is high peak-to-average power ratio (PAPR) due to the random addition
of subcarriers in the time domain. For instance, consider the four
sinusoidal signals as shown in Fig. \ref{fig:PAPR} with different
frequencies and phase shifts \cite{rahmatallah2013}. The resulting
signal envelope presents high peaks when the peak amplitudes of the
different signals are aligned at the same time. As a result of such
high peaks, the power amplifier at the transmitter operates in the
nonlinear region causing a distortion and spectral spreading. In addition,
as the number of subcarriers increases, the variance of the output
power increases as well.

\begin{figure}[t]
\centering\includegraphics[width=0.5\columnwidth]{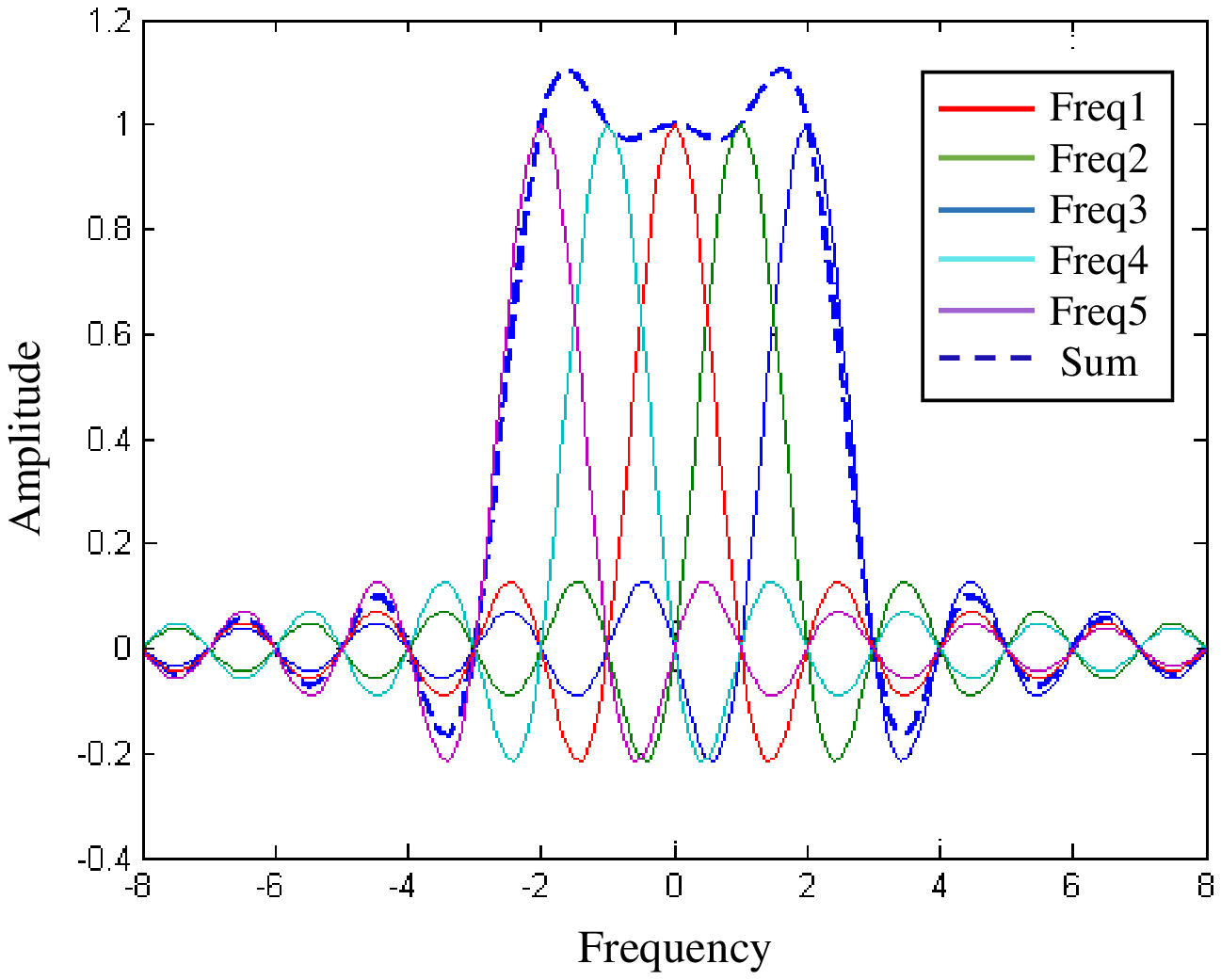}
\centering{}\caption{The OOBE problem in multicarrier schemes.\label{fig:OOBE}}
\end{figure}

Another critical issue related to the CP-OFDM systems is its high
out of band emissions (OOBE). The OFDM signal is well localized in
the time domain with a rectangular pulse shape that results in a sinc
shape in the frequency domain as shown in Fig. \ref{fig:OOBE}. Especially,
the sidelobes of the sincs at the edge carriers cause significant
interference and should be reduced to avoid adjacent channel interference
(ACI). Typically, OOBE is reduced by various windowing/filtering approaches
along with the guard band allocation \cite{demir2017} to meet the
spectral mask requirements of the various standards. 3GPP LTE standard
uses 10\% of total bandwidth as guard bands to handle this problem.
However, fixed guard allocation decreases the spectral efficiency.

Furthermore, OFDM systems are more sensitive to synchronization errors
than single-carrier systems. As an example, if the orthogonality is
lost due to the frequency offset, Doppler spread, or phase noise,
the leakage from other subcarriers causes ICI. Similarly, the timing
offset causes ISI or ICI when it occurs outside the guard interval. 

\begin{figure}[b]
\centering\includegraphics[width=0.8\columnwidth]{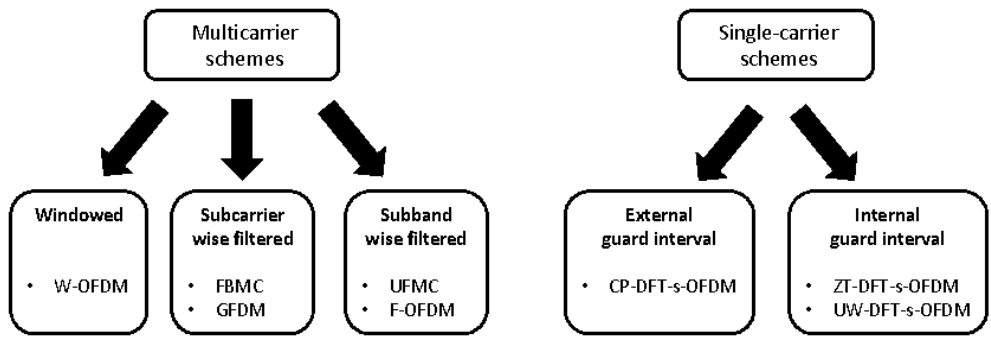}
\centering{}\caption{Major waveforms candidates for 5G and beyond.\label{fig:WaveformTree}}
\end{figure}

Numerous waveforms are proposed considering all these disadvantages
for the upcoming 5G standard \cite{3gppQualCand,zhang2016,berardinelli2016,lin2016,sahin2016}.
Although backward compatibility, low implementation complexity, and
easy multiple-input multiple output (MIMO) integration still make
CP-OFDM an important candidate for the new standards, it seriously
suffers from its limited flexibility and the unfriendly coexistence
with different numerologies for various channel conditions and use
cases in addition to the aforementioned problems. The proposed waveforms
provide better flexibility and time\textendash frequency localization
using various filtering/windowing approaches and precoding strategies
with certain trade-offs. Also, the external guard interval, that is
CP, is suggested to be replaced with flexible internal guard interval
to improve spectral efficiency further and to provide better performance.
However, it is only being considered for the single-carrier schemes
for practical reasons currently. The major waveform candidates for
5G and beyond are classified, as shown in Fig. \ref{fig:WaveformTree},
and discussed thoroughly in the following sections. 

\section{Major Waveform Candidates for 5G and Beyond \label{sec:III}}

\subsection{Multicarrier Schemes}

\subsubsection{Windowing}

\begin{figure}[t]
\centering\includegraphics[width=0.8\columnwidth]{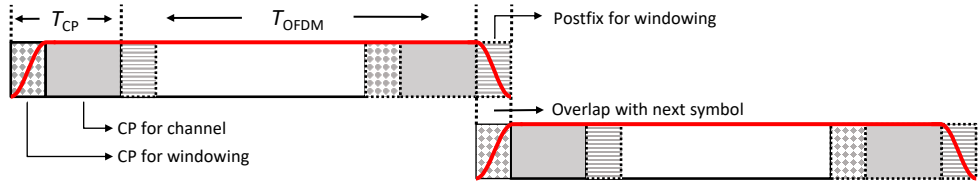}
\centering{}\caption{W-OFDM (Transmitter).\label{fig:W-OFDM}}
\end{figure}

The discontinuity between adjacent symbols due to the rectangular
window shape in the time domain causes high OOBE for CP-OFDM. Windowed-OFDM
(W-OFDM) \textbf{\cite{3gppQualCand}} smooths these sharp edges in
a straightforward way with low complexity. The baseband W-OFDM can
be expressed as follows: 

\begin{equation}
s_{W-OFDM}[k]=\sum_{m=-\infty}^{+\infty}\sum_{n=0}^{N-1}d_{m,n}g[k-m(N+L_{CP}+L_{Ext})]e^{j2\pi k\frac{n}{N}}\label{eq:Eq2-W-OFDM}
\end{equation}
where $d_{m,n}$ is the complex data transmitted on the $n$\textsuperscript{th}
subcarrier and $m$\textsuperscript{th} OFDM symbol, $L$\textsubscript{CP}
presents the CP length, $L$\textsubscript{Ext} expresses the windowing
extension, and $g[n]$ shows the windowing function. Several windowing
functions have been evaluated in detail \cite{farhang2011} with different
trade-offs between the width of the main lobe and suppression of the
side lobes. An illustration of windowing operation at the transmitter
is shown in Fig. \ref{fig:W-OFDM}.

\begin{figure}[b]
\centering\includegraphics[width=0.8\columnwidth]{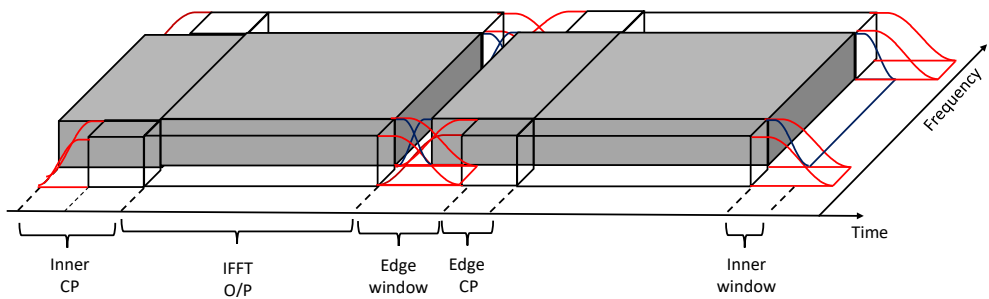}
\centering{}\caption{Edge windowing technique \cite{sahin2011}.\label{fig:Edge-Window}}
\end{figure}

\begin{figure}[t]
\centering\subfloat[]{\includegraphics[width=0.8\columnwidth]{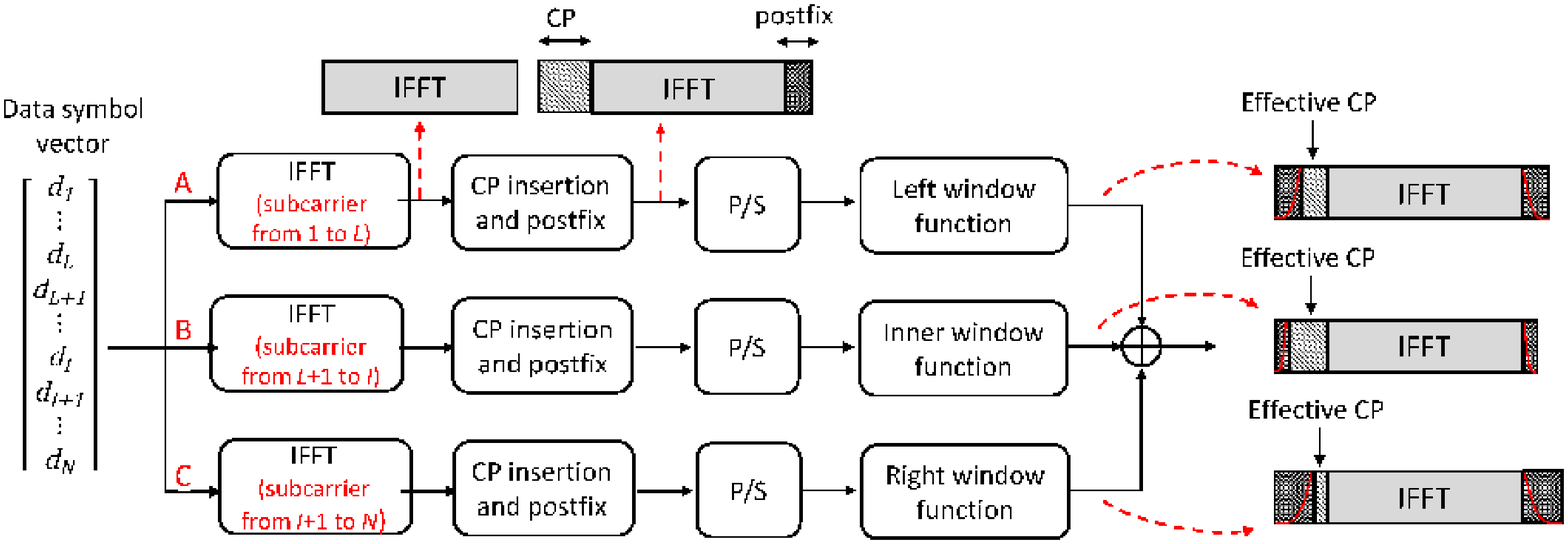}

}\\\subfloat[]{\includegraphics[width=0.8\columnwidth]{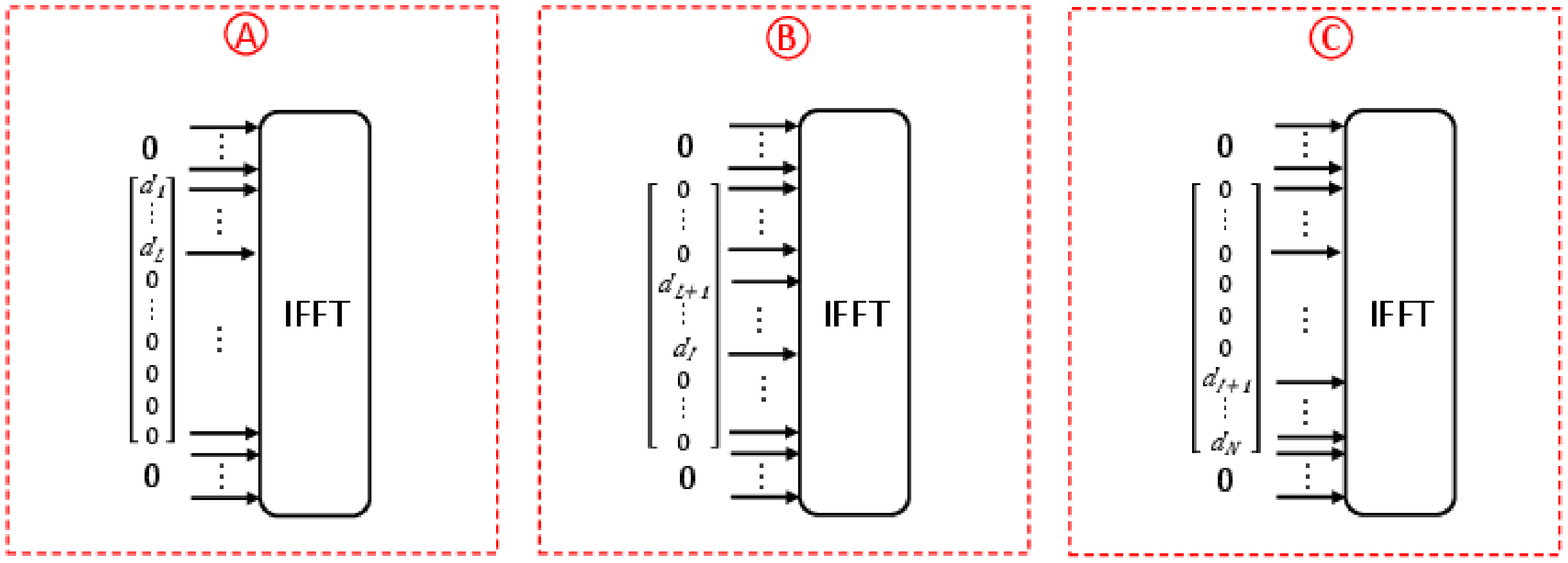}

}

\caption{Edge windowing block diagram \cite{3gppSamsungEdge}.\label{fig:Edge-Window-Block}}
\end{figure}

Initially, the CP is further extended on both edges at the transmitter
and the extended part from the beginning of the OFDM symbol is appended
to the end. The windowing operation is applied symmetrically on both
edges of the OFDM symbol, and the transitions parts (i.e., ramp-ups
and ramp-downs) of adjacent symbols are overlapped to shorten the
extra time domain overhead resulting from windowing. In addition,
the windowing operation is performed at the receiver as well to reduce
the interference from other users.

Edge windowing \cite{sahin2011} is another approach to reduce the
high OOBE of CP-OFDM. It is well known that the outer subcarriers
have a higher influence on OOBE problem compared to the inner subcarriers.
However, conventional windowing techniques apply the same window for
all subcarriers within an OFDM symbol. As a result, the spectral efficiency
decreases due to an extra windowing duration or the performance degrades
since the effective CP length of channel shortens. The proposed approach
borrows the CP duration of the channel to perform windowing and maintains
the spectral efficiency (Fig. \ref{fig:Edge-Window}). The longer
windows that decrease the effective CP size are applied only to the
edge subcarriers, as shown in Fig. \ref{fig:Edge-Window-Block} and
hence the OOBE is suppressed with a minimal performance loss. If the
CP of the edge subcarriers is less than or equal to the maximum excess
delay of the channel, neither ISI nor ICI is observed. Therefore,
these edge subcarriers should be assigned to the user equipments that
experience shorter delay spread. The edge windowed OFDM provide a
better spectrum confinement with a low complexity and negligible performance
loss \cite{3gppSamsungEdge}.

Although windowing approaches present lower OOBE compared to CP-OFDM,
the effect is limited and nonnegligible guard bands are still required.
However, these methods can be applied along with the filtering approaches
that are discussed in the following sections to provide better spectral
confinement. 

\medskip{}

\subsubsection{Subcarrier-Wise Filtering}
\begin{description}
\item [{a)}] FBMC
\end{description}

\begin{figure}[t]
\centering\includegraphics[width=0.75\columnwidth]{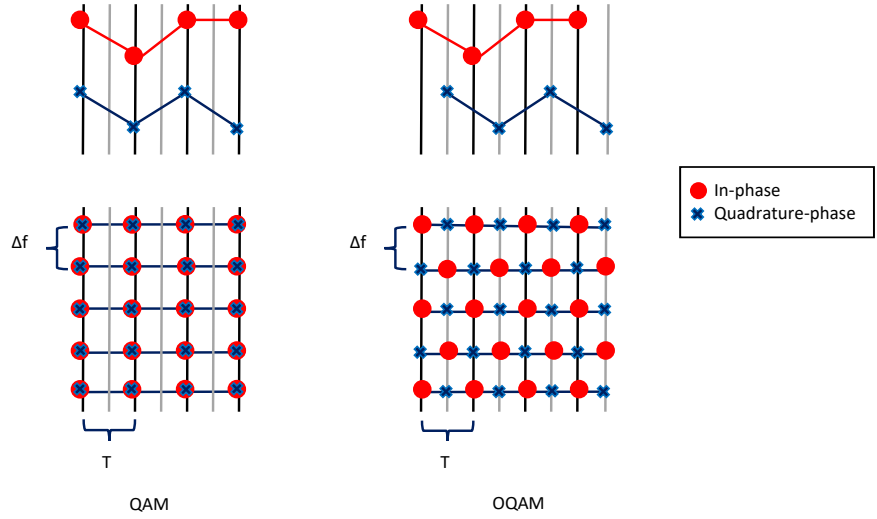}
\centering{}\caption{QAM signaling versus OQAM signaling \cite{tech5gRS}. \label{fig:OQAM-FBMC}}
\end{figure}

Filter bank multicarrier (FBMC) yields a good frequency-domain localization
by extending the pulse duration in the time domain and using properly
designed pulse shaping filters \cite{sahin2014,farhang2011}. These
flexible filters are applied at the subcarrier level, and they enable
adaption to various channel conditions and use cases. There are various
ways to implement FBMC such as filtered multitone (FMT), cosine modulated
multitone (CMT), and staggered modulated multitone (SMT). However,
SMT, which is widely known as offset quadrature amplitude modulation
OQAM\textendash FBMC, is the focus of 5G waveform discussion due to
its ability to handle interference while allowing dense symbol placement
in the time\textendash frequency lattice \cite{zhang2016}. OQAM signaling
provides staggering of \textquotedblleft in-phase\textquotedblright{}
and \textquotedblleft quadrature-phase\textquotedblright{} components
in both time and frequency domains, as shown in Fig. \ref{fig:OQAM-FBMC}
and hence, orthogonality is maintained within the real and imaginary
domains separately.

\begin{figure}[b]
\centering\includegraphics[width=0.8\columnwidth]{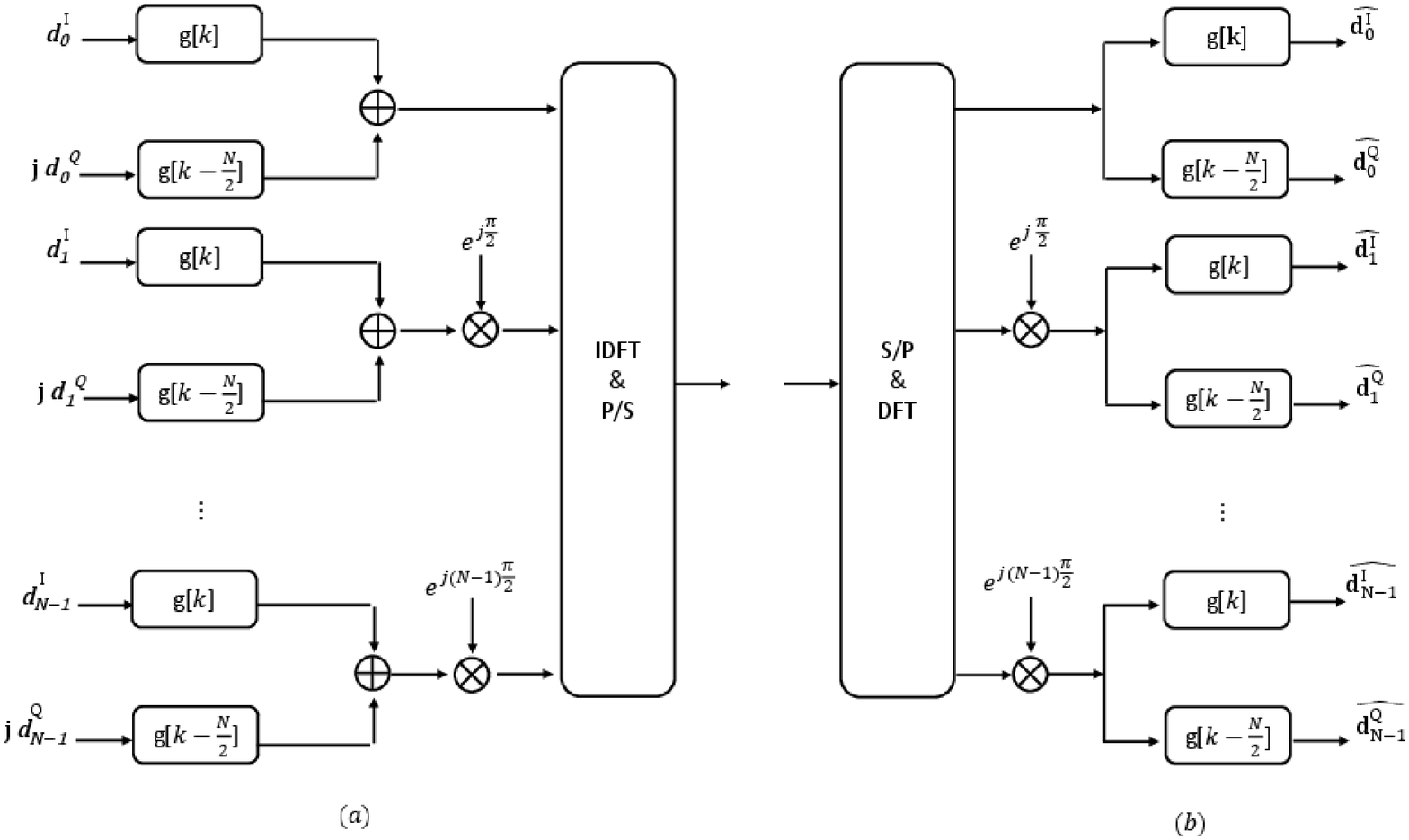}
\centering{}\caption{OQAM-FBMC block diagram. (a) Transmitter. (b) Receiver.\label{fig:OQAM-FBMC-BlockDiagram} }
\end{figure}

The baseband FBMC/OQAM is expressed as follows:

\begin{equation}
s_{FBMC-OQAM}[k]=\sum_{m=-\infty}^{+\infty}\sum_{n=0}^{N-1}d_{m,n}g[k-m\frac{N}{2}]e^{j2\pi k\frac{n}{N}}e^{j\phi_{m,n}}\label{eq:Eq3-FBMC}
\end{equation}
where $g$ represents the prototype filter, $\phi_{m,n}$ is an additional
phase term at subcarrier $n$ and symbol index $m$, which is expressed
as $\frac{\pi}{2}(m+n)$. The $d_{m,n}$ is real valued since the
real and imaginary parts are transmitted with a delay. Also, to address
a perfect reconstruction of symbols, the prototype filter must satisfy
the orthogonality condition \cite{lin2016}. A block diagram of conventional
OQAM\textendash FBMC transmitter and receiver is shown in Fig. \ref{fig:OQAM-FBMC-BlockDiagram}.

The subcarriers are well localized in the frequency domain due to
the utilization of prototype filters and are spread over only a few
subcarriers in FBMC systems. Furthermore, the orthogonality between
neighbor subcarriers is ensured using OQAM. As a result, the equalization
is simplified without the use of CP, and no more than one subcarrier
is required as a guard band for non-orthogonal transmissions \cite{zhang2016}.
The savings on both guard band and guard duration enable this waveform
to achieve better spectrum efficiency compared to CP-OFDM. Also, the
well-localized subcarriers in OQAM\textendash FBMC make it suitable
for high mobility applications as it is more immune to the Doppler
effects. On the other hand, there exist several practical challenges
currently. The MIMO integration and pilot design with OQAM\textendash FBMC
are not straightforward as in CP-OFDM due to the intrinsic interference
resulting from OQAM signaling \cite{lin2015}. 
\begin{description}
\item [{b)}] GFDM
\end{description}

\begin{figure}[b]
\centering\includegraphics[width=0.9\columnwidth]{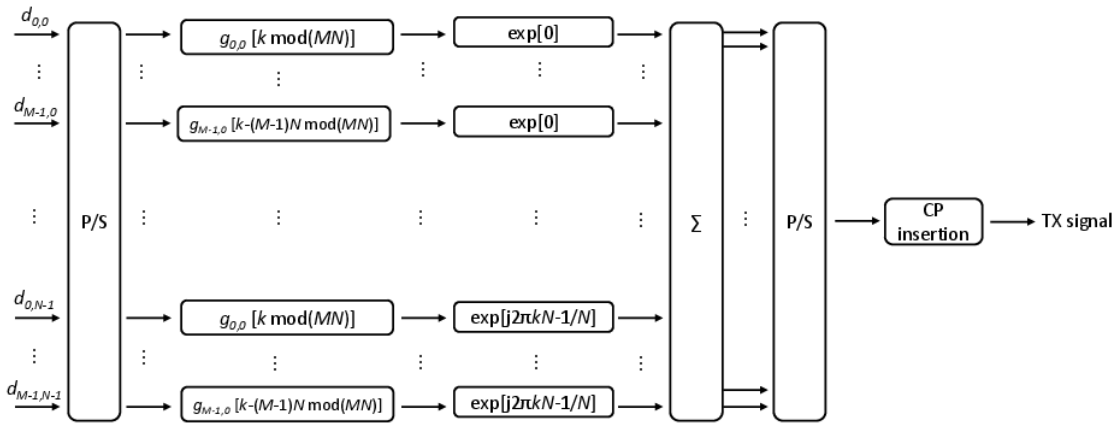}
\centering{}\caption{GFDM block diagram (Transmitter).\label{fig:GFDM-Block}}
\end{figure}

Generalized frequency division multiplexing (GFDM) \cite{michailow2014}
also applies subcarrier filtering similar to FBMC. However, the filters
for pulse shaping are circularly convoluted over a defined number
of symbols. Also, the symbols are processed blockwise, and CP is appended
to this block. Considering $N$ subcarriers in each subsymbol group
and $M$ subsymbol group in each block, a GFDM symbol is represented
as follows: 

\begin{equation}
s_{GFDM}[k]=\sum_{m=0}^{M-1}\sum_{n=0}^{N-1}d_{m,n}g_{m,n}[(k-mN)mod(MN)]e^{j2\pi k\frac{n}{N}}\label{eq:Eq4-GFDM}
\end{equation}
where $d_{m,n}$ is the complex data transmitted on the $n$\textsuperscript{th}
subcarrier and $m$\textsuperscript{th} sub symbol, and $g[n]$ shows
the prototype filter. Although FBMC prototype filters must satisfy
orthogonality condition, there is no constraint on GFDM prototype
filters \cite{lin2016}. Hence, GFDM is usually a nonorthogonal transmission
scheme with nonorthogonal filters. A block diagram of conventional
GFDM transmitter is shown in Fig. \ref{fig:GFDM-Block}. 

\begin{figure}[t]
\centering\includegraphics[width=0.9\columnwidth]{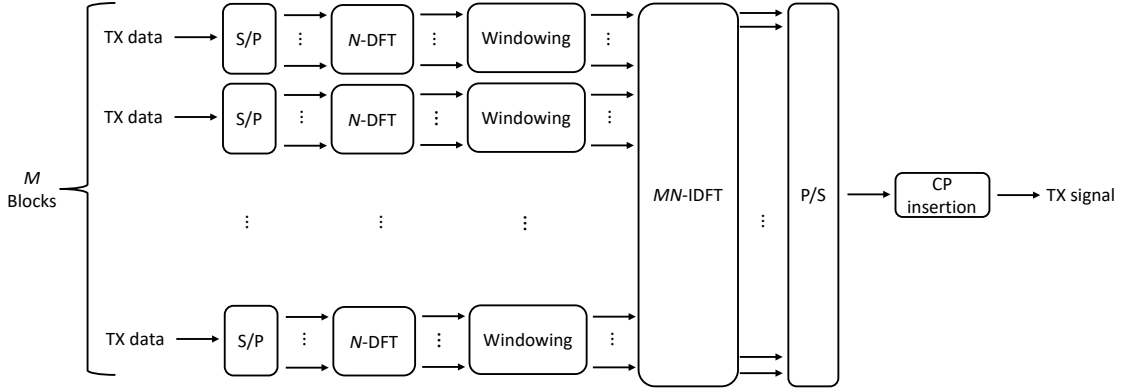}
\centering{}\caption{Equivalency of GFDM and DFT-s-OFDM (Transmitter).\label{fig:GFDM-Block-2}}
\end{figure}

GFDM is proposed as a flexible waveform where the number of subsymbols,
subcarriers, and prototype filters are adjustable for various channel
conditions and use cases. Conceptually, a GFDM signal can also be
generated with $M$ FFTs of size $N$, filter banks, and an $MN$-point
IFFT, as shown in Fig. \ref{fig:GFDM-Block-2}. From this implementation
perspective, it is equivalent to a DFT-s-OFDM signal when the rectangular
function is used as a prototype filter, which also explains lower
PAPR compared to CP-OFDM. This equivalency is further discussed in
the following single-carrier waveform discussion. Also, it is equivalent
to CP-OFDM when $M$ equals to 1. 

GFDM shares the well-frequency-localized characteristic with OQAM-FBMC.
Hence, it is suitable for high-mobile scenarios and provides more
immunity to synchronization errors. Although GFDM provides flexibility
in the waveform, the nonorthogonal transmission scheme requires complex
successive interference cancellation (SIC) algorithms at the receiver.
Similar to OQAM-FBMC, pilot design and MIMO transmission is complicated.
Furthermore, the block-wise transmission causes latency that makes
it infeasible for mission-critical applications.

The waveforms that perform subcarrier-wise filtering, that is FBMC
and GFDM require a new transceiver design, and there is no backward
compatibility with 4G LTE. The next section describes the subband-wise
filtered multicarrier schemes and these operations are applicable
to the current standards without significant changes. 

\subsubsection{Subband-Wise Filtered MCM}
\begin{description}
\item [{a)}] UFMC
\end{description}

\begin{figure}[t]
\centering\includegraphics[width=0.9\columnwidth]{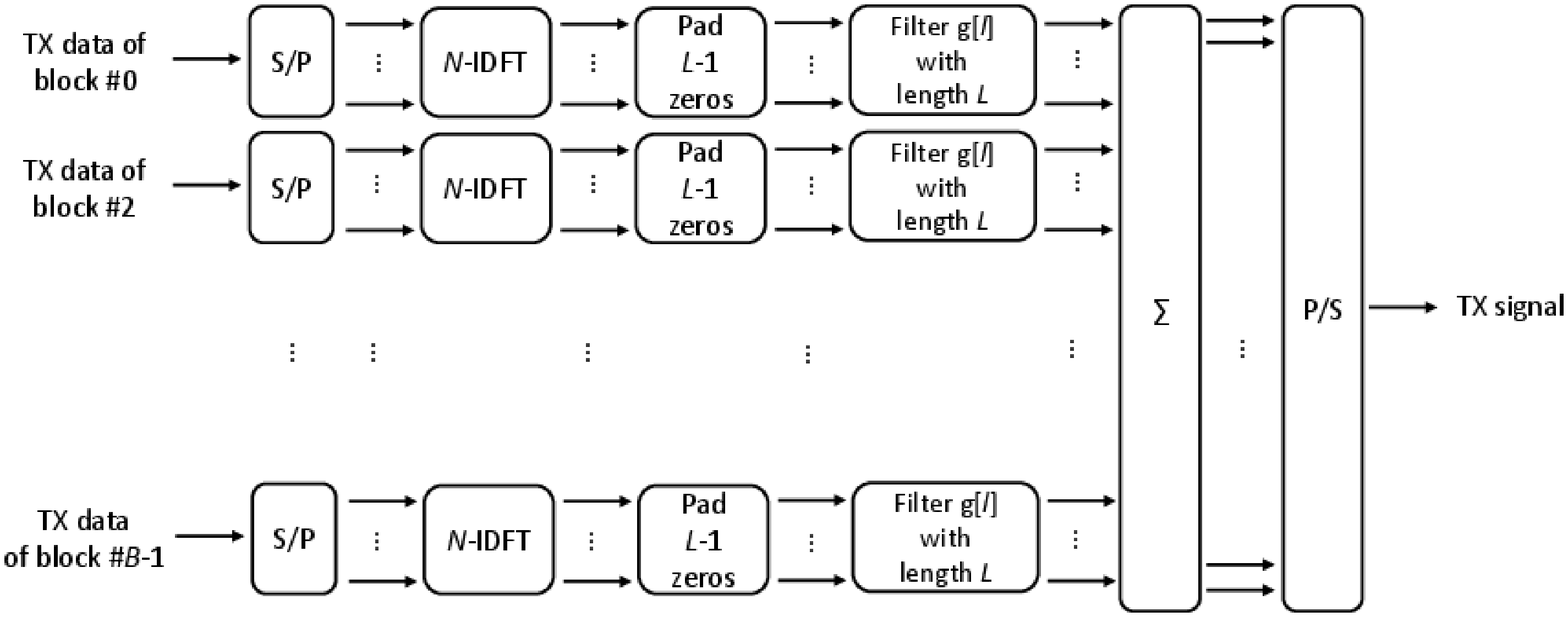}
\centering{}\caption{UFMC block diagram (Transmitter).\label{fig:UFMC-Block}}
\end{figure}

Universal filtered multicarrier (UFMC) \cite{vakilian2013} applies
subband-wise filtering to reduce OOBE. The subband-wise filtering
is considered as a compromise between the whole band filtering and
subcarrier-wise filtering. Hence, the filters are shorter compared
to the FBMC, where the length of filters are much longer than the
symbol duration. The total available bandwidth is partitioned into
subbands and filtering is performed with a fixed frequency-domain
granularity \cite{zhang2016}. Considering $B$ subbands (blocks)
in total and using a filter of length $L$, the baseband UFMC signal
is represented as follows:

\begin{equation}
s_{UFMC}[k]=\sum_{b=0}^{B-1}\sum_{l=0}^{L-1}\sum_{n=0}^{N-1}d_{n}^{b}g[l]e^{j2\pi k\frac{(n-l)}{N}}\label{eq:Eq5-UFMC}
\end{equation}
where $d_{n}^{b}$ is the complex data transmitted on the $n$\textsuperscript{th}
subcarrier and $b$\textsuperscript{th} subband, and $g[l]$ shows
the frequency equivalent windowing function of a time domain finite
impulse response (FIR) filter. Therefore, each block length is $L+N-1$.
The use of CP is optional to provide better immunity against ISI,
and it is also called as UF-OFDM when CP is used. However, typical
UFMC systems do not utilizes CP and the transitions regions (i.e.,
ramp-ups and ramp-downs) provide a soft ISI protection. A block diagram
of conventional UFMC transmitter is shown in Fig. \ref{fig:UFMC-Block}.

The symbols are sent back-to-back without any overlapping, and hence
orthogonality in time is maintained. However, the symbols are not
circularly convoluted with the channel due to lack of CP and a more
complicated receiver is required \cite{schaich2014}. A conventional
UFMC receiver utilizes an FFT block that has twice the size of IFFT
block at the UFMC transmitter.

UFMC provides a better localization in the frequency domain and robustness
against time\textendash frequency offsets compared to CP-OFDM. Also,
shorter filter lengths compared to subcarrier-wise filtering makes
it more suitable for low-latency applications. On the other hand,
these shorter filters offer limited OOBE suppression. Furthermore,
increased complexity due to the lack of CP and complicated filtering
operations should be dealt with intelligently to design practical
communications systems.
\begin{description}
\item [{b)}] f-OFDM
\end{description}

\begin{figure}[t]
\centering\includegraphics[width=0.63\columnwidth]{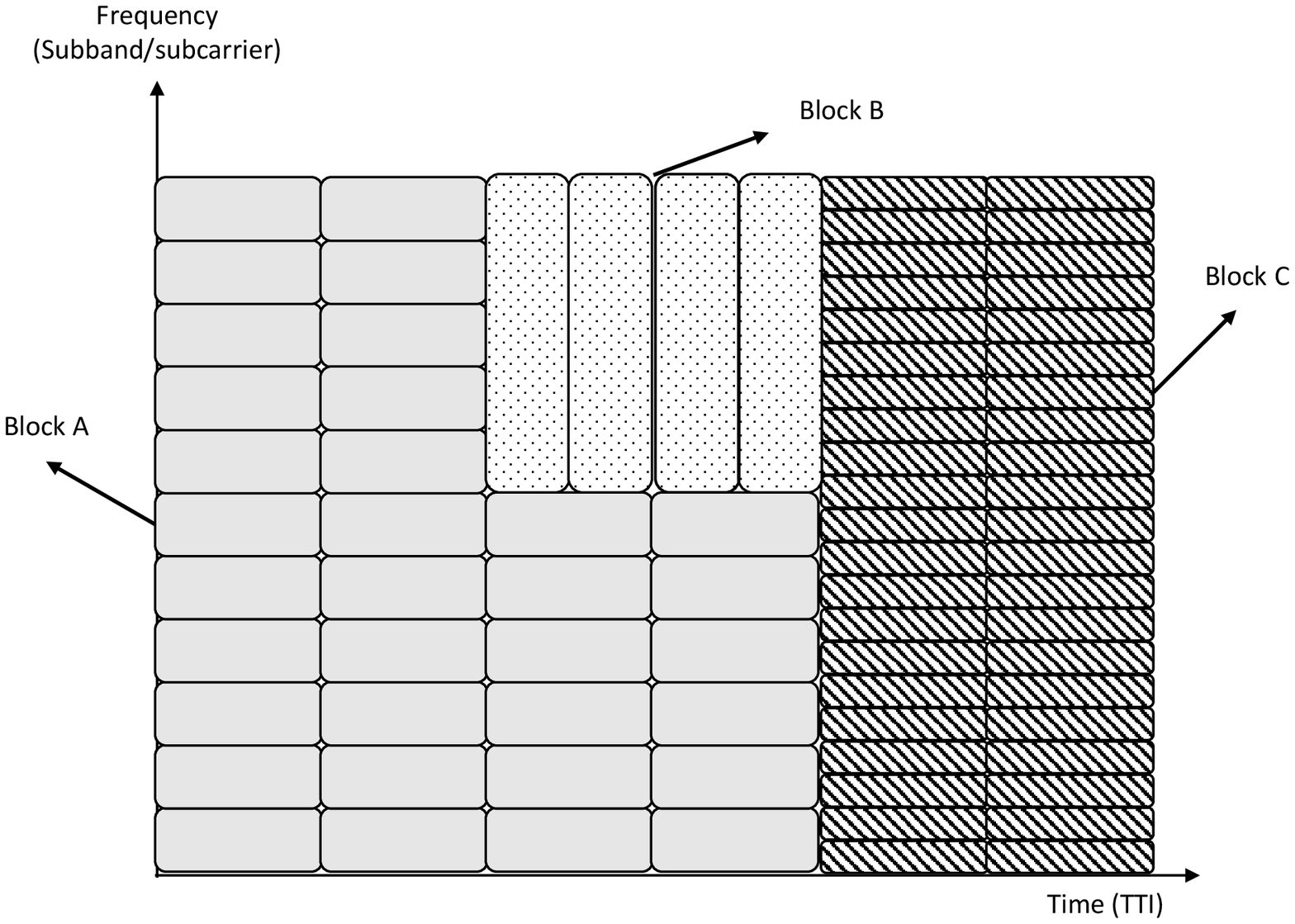}
\centering{}\caption{An example of flexible resource allocation enabled by f-OFDM.\label{fig:f-OFDM-Lattice}}
\end{figure}

Filtered-OFDM (f-OFDM) \cite{zhang2015} is another subband-wise filtered
multicarrier scheme, but the filtering granularity is more flexible
than UFMC. The partition in the time\textendash frequency grid is
adjusted based on the different channel conditions and use cases,
as shown in Fig. \ref{fig:f-OFDM-Lattice}. This flexibility makes
f-OFDM more suitable for the use of different numerologies (such as
bandwidth, sub-carrier spacing, CP duration, and transmission time
interval) compared to UFMC with the cost of increased complexity.
Considering $B$ blocks in total, the baseband f-OFDM signal is represented
as follows:

\begin{equation}
s_{f-OFDM}[k]=\sum_{b=0}^{B-1}\sum_{m=0}^{M-1}\sum_{l=0}^{L_{b}-1}\sum_{n=0}^{N-1}d_{m,n}^{b}g_{b}[l]e^{j2\pi k\frac{(n-l-mL_{CP})}{N}}\label{eq:Eq6-F-OFDM}
\end{equation}
where $d_{m,n}^{b}$ is the complex data transmitted on the $b$\textsuperscript{th}
block, $n$\textsuperscript{th} subcarrier, and $m$\textsuperscript{th}
subsymbol, $g_{b}[l]$ shows the frequency equivalent windowing function
of a time domain FIR filter on the $b$\textsuperscript{th} block,
and $L$\textsubscript{CP} presents the CP size. 
\begin{figure}[b]
\centering\includegraphics[width=0.9\columnwidth]{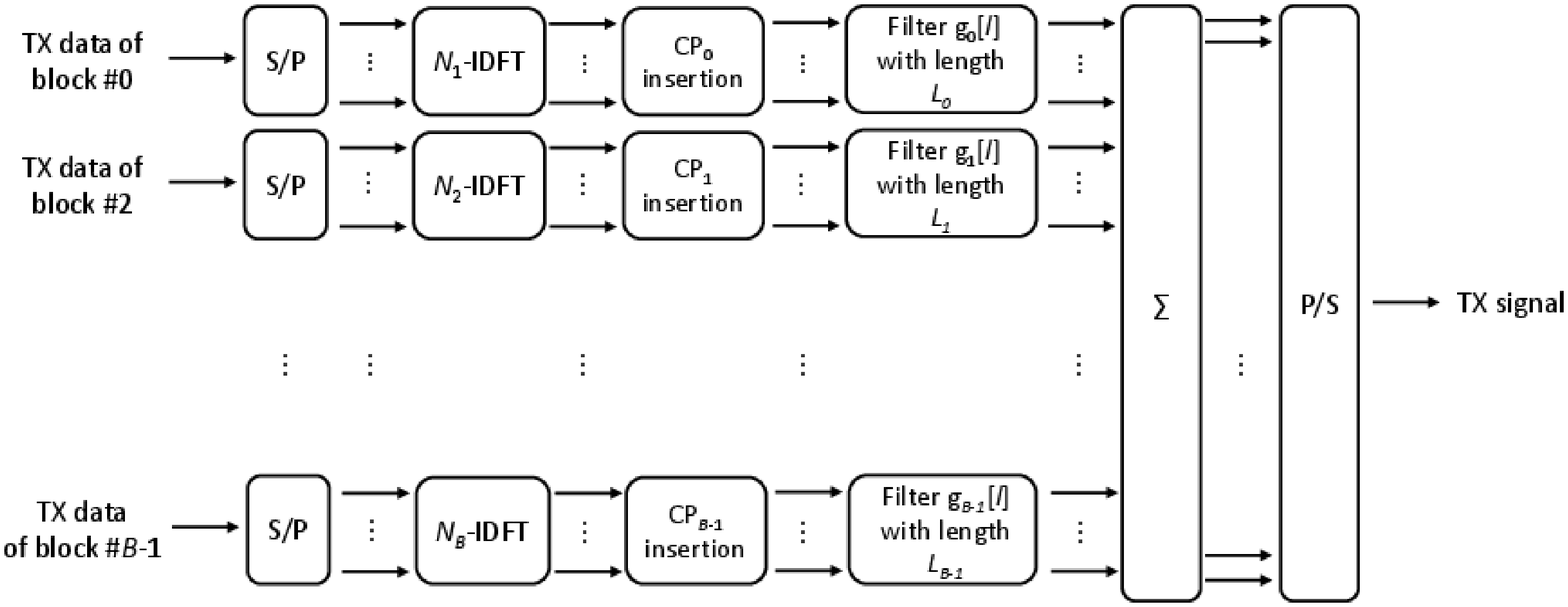}
\centering{}\caption{f-OFDM block diagram (Transmitter).\label{fig:f-OFDM-Block} }
\end{figure}
As can be seen in the equation, f-OFDM maintains the CP in contrast
to UFMC. Therefore, it is more immune to the ISI and needs less complex
receiver. Ideally, the frequency domain window $g_{b}[l]$ is desired
to be a rectangle with a size of $L_{b}$. However, it corresponds
to an infinite length sinc shape response in the time domain, and
hence it is impractical. Therefore, windowed sinc functions are used
in the filtering operation. More details on various filters can be
found in Ref. \cite{farhang2011}. A block diagram of conventional
f-OFDM transmitter is shown in Fig. \ref{fig:f-OFDM-Block}. Matched
filtering and identically sized IFFT/FFT blocks at the receiver also
differs f-OFDM from the other subband-wise filtering technique, UFMC. 

F-OFDM shares all advantages of well-frequency-localized waveforms
such as low OOBE, allowing asynchronous transmission, being feasible
for different numerologies, and requiring less number of guard tones.
Although f-OFDM cannot provide low OOBE as subcarrier-wise filtered
multicarrier schemes due to the use of shorter filter lengths, it
is compatible with MIMO transmission scheme and does not require any
successive interference cancellation (SIC) algorithm. However, complexity
is still the main drawback of f-OFDM compared to CP-OFDM. 

\subsection{Single-Carrier Schemes}
\begin{description}
\item [{a)}] CP-DFT-s-OFDM
\end{description}

\begin{figure}[b]
\centering\includegraphics[width=0.8\columnwidth]{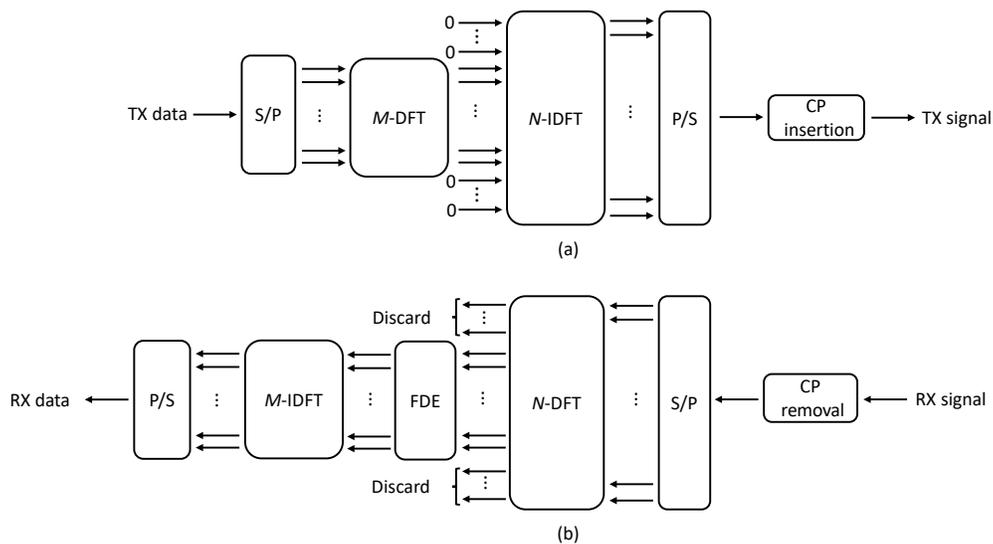}
\centering{}\caption{CP-DFT-s-OFDM block diagram. (a) Transmitter. (b) Receiver.\label{fig:DFT-s-OFDM-Block}}
\end{figure}

Discrete Fourier transform spread OFDM (DFT-s-OFDM) waveforms are
proposed to mitigate the high PAPR problem in CP-OFDM while maintaining
the useful characteristics of it. The data input can be modeled as
independent and identically distributed random variables, and as a
result, the corresponding output of IDFT in CP-OFDM has a high variance.
Such high variance is reduced by providing correlation to the input
data by performing a DFT operation before the IDFT process, as shown
in Fig. \ref{fig:DFT-s-OFDM-Block}.

The utilization of CP in DFT-s-OFDM ensures circularity of the signal
at the receiver and enables easy FDE to handle multipath channel effect.
This waveform can be interpreted in two ways \cite{sahin2016}. One
interpretation is that it is a precoded CP-OFDM scheme, where PAPR
is mitigated by DFT precoding. This interpretation provides to consider
different precoding methods. The other interpretation is that it is
a transmission scheme that upsamples the input data by the ratio of
the IDFT and DFT block sizes (i.e., $N/M$ where $N>M$), and performs
a circular pulse shaping with a Dirichlet sinc function. This interpretation
lets the designers consider different pulse shaping approaches to
reduce the PAPR further and handle the OOBE.

DFT-s-OFDM is deployed in the uplink of the 4G LTE due to its lower
PAPR feature that provides better power efficiency. The low complexity,
support of dynamic spectrum access, and MIMO compatibility make CP-DFT-s-OFDM
a significant candidate for the 5G, similar to the CP-OFDM. However,
the spectral efficiency of this waveform is also comparable to CP-OFDM
and suffers from high OOBE because of the discontinuity between adjacent
symbols. Hence, similar windowing and filtering approaches as discussed
in the multicarrier schemes discussion that can be applied to improve
spectral efficiency. In addition to that, internal guard concept for
DFT-s-OFDM is being discussed for the 5G, which provides more flexibility
in the system design as explained in the following sections. 
\begin{description}
\item [{b)}] ZT-DFT-s-OFDM
\end{description}

\begin{figure}[b]
\centering\includegraphics[width=1\columnwidth]{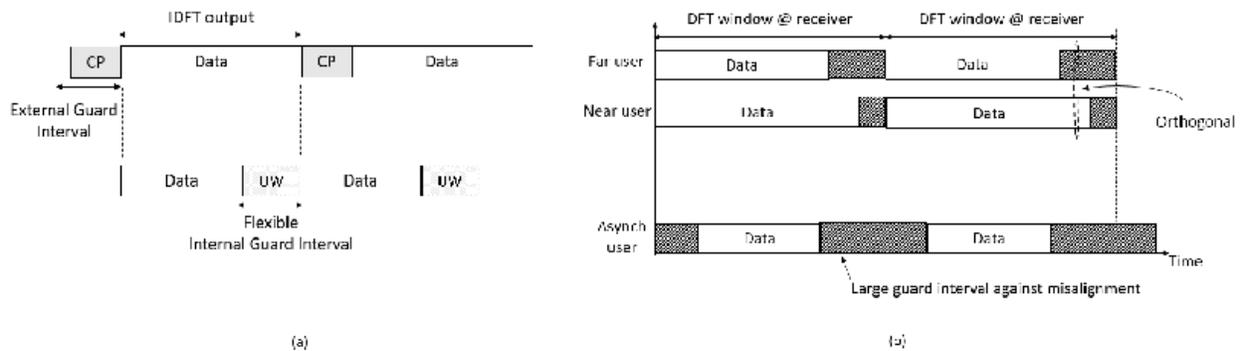}

\caption{(a) Flexible internal guard interval. (b) An illustrative example
of utilizing flexible internal guard interval in the uplink\cite{sahin2016}.\label{fig:Internal_Guard}}
\end{figure}

The guard interval is hard-coded in 4G LTE systems, and there exist
only two options as normal and extended CP. However, the base station
is preset to only one of these guard intervals because the use of
different guard interval durations results in different symbol durations
and consequently a different number of symbols per frame. This leads
to the generation of mutual asynchronous interference even when the
frames are aligned \cite{berardinelli2016}. Hence, the users with
two different CP durations do not coexist together in the same cell.
As a result, the nonflexible guard interval penalizes the user equipments
that experience better channel conditions. Zero-tail-DFT-spread-OFDM
(ZT\textendash DFT-s-OFDM) \cite{berardinelli2013} is proposed to
solve this problem. The CP is replaced with an internal guard period
that provides the same functionality. The total period of the guard
duration and data duration is fixed, but the ratio between them is
flexible as shown in Fig. \ref{fig:Internal_Guard}. The flexibility
provides better spectral efficiency while maintaining the total symbol
duration.

\begin{figure}[t]
\centering\includegraphics[width=0.7\columnwidth]{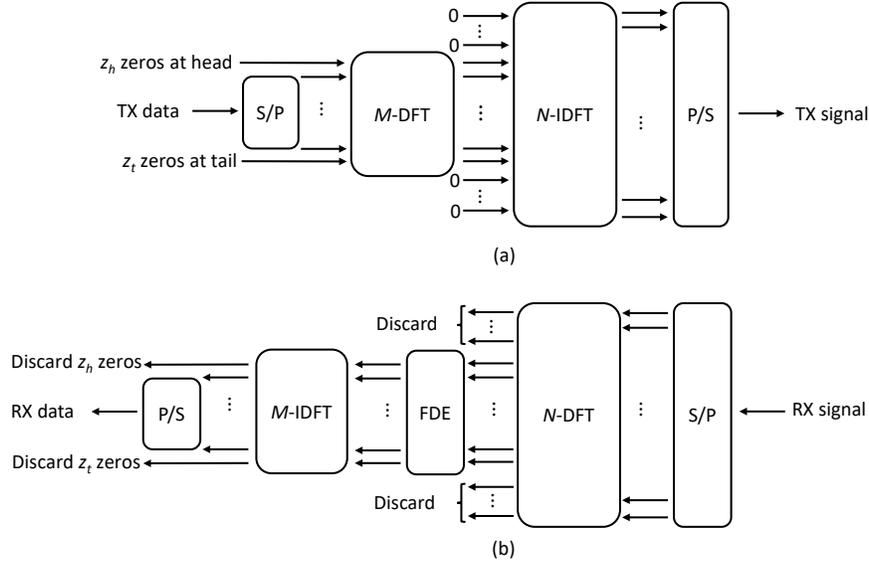}
\centering{}\caption{ZT-DFT-s-OFDM block diagram. (a) Transmitter. (b) Receiver.\label{fig:ZT-DFT-s-OFDM-Block} }
\end{figure}

Zero vectors with variable lengths are inserted into the head and
tail of the data before DFT process in this approach. The tail length
is set to be longer than the delay spread of the channel and, hence
the leakage to the next symbol does not have significant power. Also,
the zeros in the head, which are usually shorter than the zeros at
the tail, provide a smoother transition and yield substantial OOBE
reduction \cite{berardinelli2016}. A block diagram of conventional
ZT-DFT-s-OFDM transmitter and receiver is shown in Fig. \ref{fig:ZT-DFT-s-OFDM-Block}. 

The fixed sequences (i.e., zero vectors) appended to each symbol ensure
circularity at the receiver, and hence ZT-DFT-s-OFDM supports single-tap
FDE. However, a residual energy of the data part in the last samples
introduces a noncyclical leakage to the next symbol \cite{berardinelli2016}
and hence, internal guard interval approach do not provide perfect
circularity as CP does. Furthermore, this leakage is a limiting factor
in the link performance for the users utilizing high-order modulations
in a multipath environment. 

The PAPR and OOBE are low for ZT-DFT-s-OFDM, and spectral efficiency
is increased due to flexible guard interval. The internal guard feature
makes it suitable for different symbol durations without introducing
mutual asynchronous interference. However, this flexibility causes
extra overhead to track delay spread of the channel. Also, windowing
can easily be applied to decrease OOBE without further extra guard
duration.

\begin{description}
\item [{c)}] UW-DFT-s-OFDM
\end{description}

\begin{figure}[t]
\centering\includegraphics[width=0.7\columnwidth]{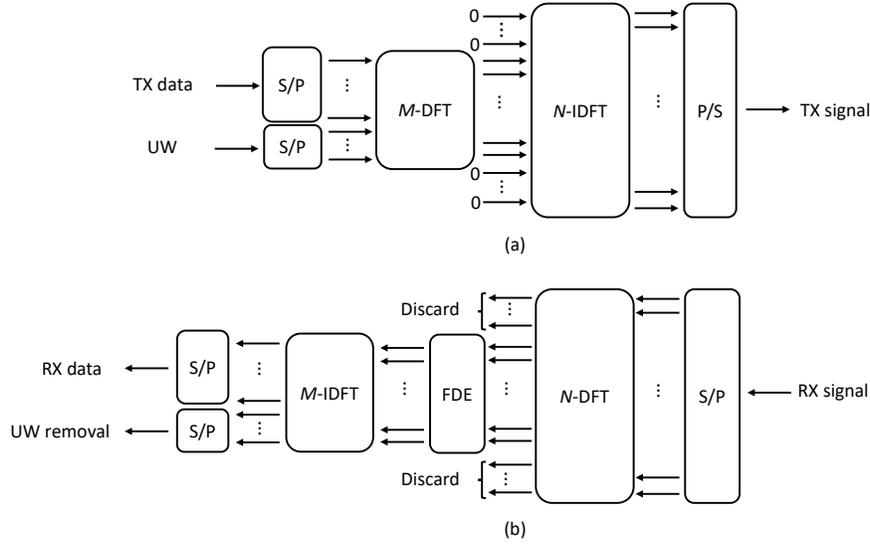}
\centering{}\caption{UW-DFT-s-OFDM block diagram. (a) Transmitter. (b) Receiver.\label{fig:UW-DFT-s-OFDM-Block} }
\end{figure}

Unique word DFT-spread OFDM (UW-DFT-s-OFDM) \cite{sahin2015} is another
single-carrier scheme that utilizes flexible internal guard band.
The zero tails and heads of the ZT\textendash DFT-s-OFDM are replaced
with a fixed sequence that enhances cyclic properties of the signal
in this approach. Since the fixed sequence is inserted before the
DFT process as shown in Fig. \ref{fig:UW-DFT-s-OFDM-Block}, the orthogonality
is provided between the data and the unique word. The circularity
is also ensured in this waveform and as a result simple FDE is supported.
However, the leakage from the data part limits its link performance
for the high-order modulations similar to ZT-DFT-s-OFDM. 

Different from the ZT part, UW can also be exploited for synchronization
and channel tracking purposes \cite{coon2006}. Therefore, it improves
spectral efficiency. The OOBE leakage characteristics of this waveform
are comparable to ZT\textendash DFT-s-OFDM since the continuity of
the symbols are provided with the DFT block. Adding data-dependent
\textquotedblleft perturbation\textquotedblright{} signal or modifying
the kernel function with windowing the input data (which is analogous
to GFDM with UW) mitigates the OOBE and PAPR further \cite{sahin2016}.
However, these benefits come with increased complexity. 

\section{Summary \label{sec:IV}}

The frequency localization is important to allow asynchronous transmission
across adjacent subbands and coexistence with other waveforms. On
the other hand, the time localization is critical for low latency
applications where longer filter/window durations are not feasible
for URLLC. All discussed waveforms for 5G provide lower OOBE compared
to CP-OFDM and its single-carrier equivalent, that is, CP-DFT-s-OFDM.
The subcarrier-wise filtering operation in FBMC results in the best
frequency localization among the candidate waveforms due to the use
of longer filter lengths. Although GFDM is another subcarrier-wise
filtered waveform, the rectangular window shape in the time domain
causes abrupt transitions and increases OOBE. However, windowing can
be performed on this waveform, and W-GFDM presents a good spectral
confinement as well. Furthermore, relatively shorter filters in the
subband-wise filtered waveforms lead to a better time localization
with a price of increasing the OOBE compared to the subcarrier-wise
filtered waveforms.

Most multicarrier schemes suffer from high PAPR and are not suitable
when high energy efficiency is required. However, GFDM exhibits a
reduced PAPR characteristic due to its equivalency to DFT-spread waveforms,
as discussed before. The single-carrier schemes are preferable in
energy-limited use cases along with the use of flexible guard intervals
that provide better spectral confinement and improved PAPR.

The spectral efficiency is another critical design criteria that is
highly affected by the window/filter duration, the shape of filter,
and extra overheads. Well-frequency-localized waveforms reduce the
need for guard bands and hence leading to better efficiency in the
frequency domain. On the other hand, the waveforms that do not utilize
a guard interval, such as FBMC, are expected to have higher efficiency
in the time domain. However, BER/BLER performance decreases substantially
in a multipath fading channel due to lack of guard interval. As a
result, complex receivers are required since an easy FDE is not possible.
MIMO compatibility is also essential to achieve high throughput. The
schemes that allow interference, such as FBMC and GFDM, cannot deploy
straightforward MIMO algorithms. 

Finally, the guard interval in the time domain makes a waveform more
robust against ISI and time-offsets. In addition, the guard bands
or the use of well-localized waveforms in the frequency domain make
a waveform robust against carrier frequency offset and Doppler effects
that reduce ICI and adjacent channel interference (ACI) in a multiple
access environment. As a result, FBMC has the best immunity to ICI
and is the most vulnerable to ISI.

\begin{figure}[H]
\centering\includegraphics[width=0.85\columnwidth]{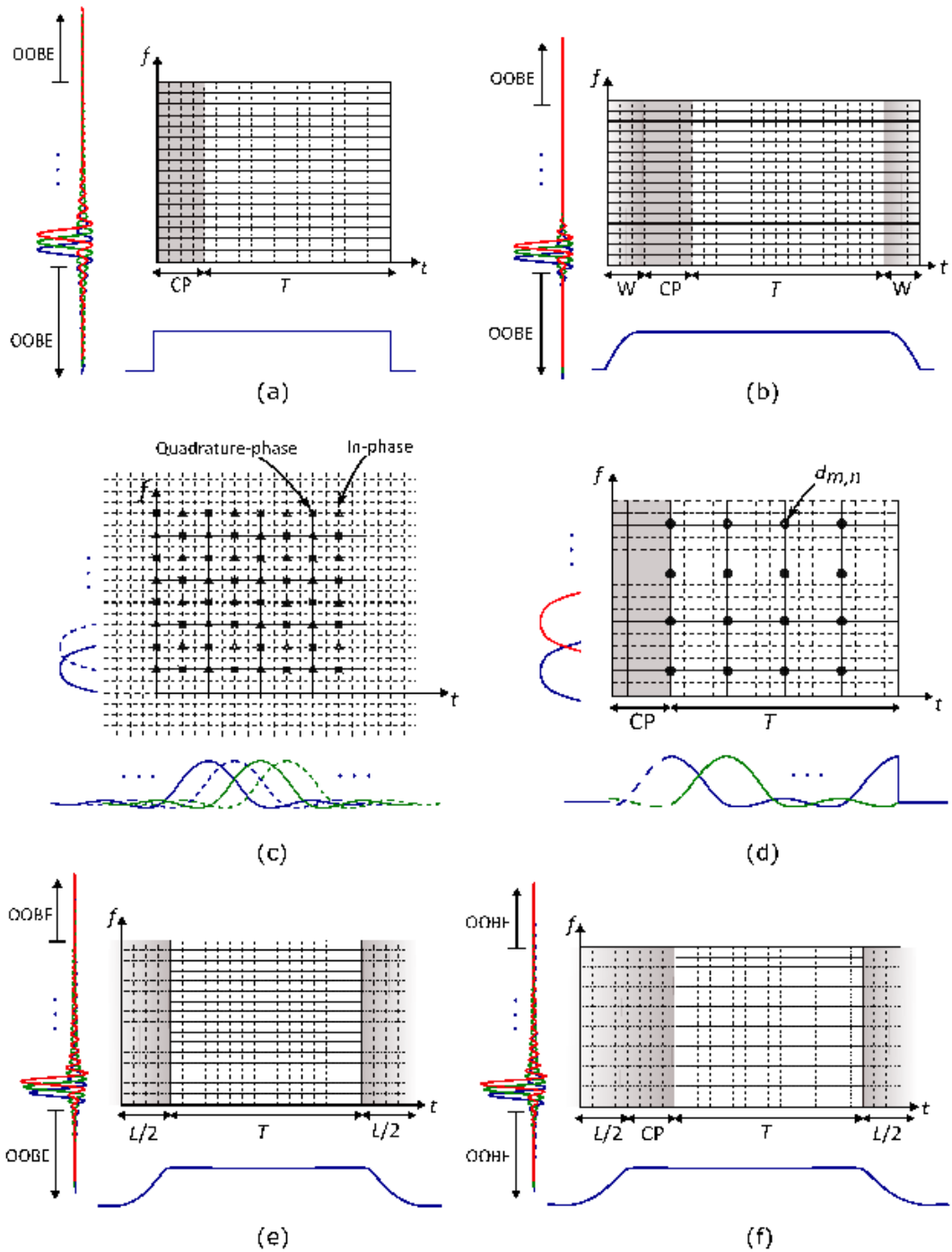}
\centering{}\caption{Comparison of multicarrier schemes. (a) CP-OFDM, (b) W-OFDM, (c) OQAM-FBMC,
(d) GFDM, (e) UFMC, and (f) f-OFDM.\label{fig:CompMC}}
\end{figure}

\begin{figure}[H]
\centering\includegraphics[width=0.85\columnwidth]{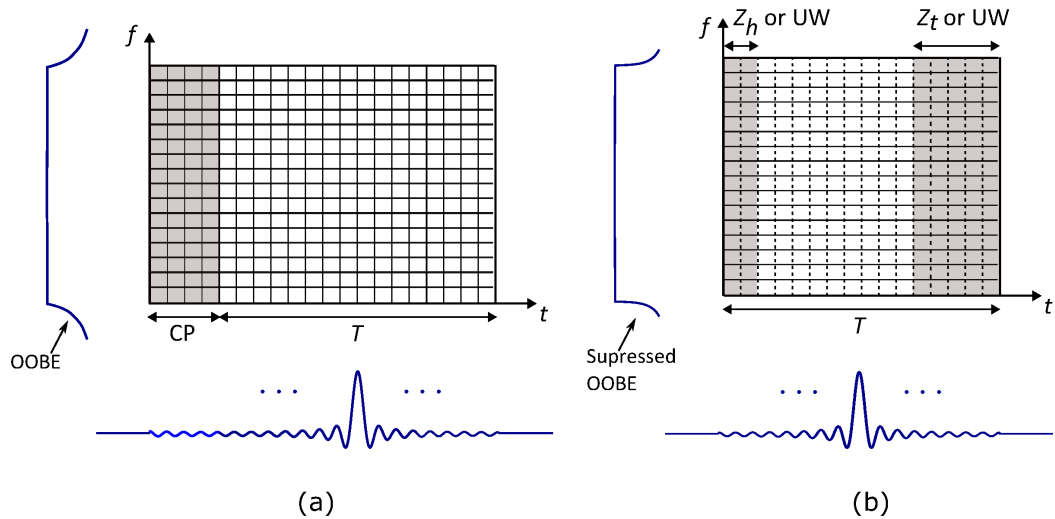}
\centering{}\caption{Comparison of single-carrier schemes. (a) CP-DFT-s-OFDM. (b) ZT-DFT-s-OFDM
or UW-DFT-s-OFDM.\label{fig:CompSC} }
\end{figure}

A summary of the main advantages/disadvantages of these major 5G candidate
waveforms is provided in Table \ref{tab:CompTable}. Moreover, the
time\textendash frequency grids, OOBE, and the pulse shapes of these
waveforms are presented in Fig. \ref{fig:CompMC} and Fig. \ref{fig:CompSC}.

\begin{table}[H]
\caption{The 5G waveform candidates. \label{tab:CompTable}}

\centering\includegraphics[width=0.85\columnwidth]{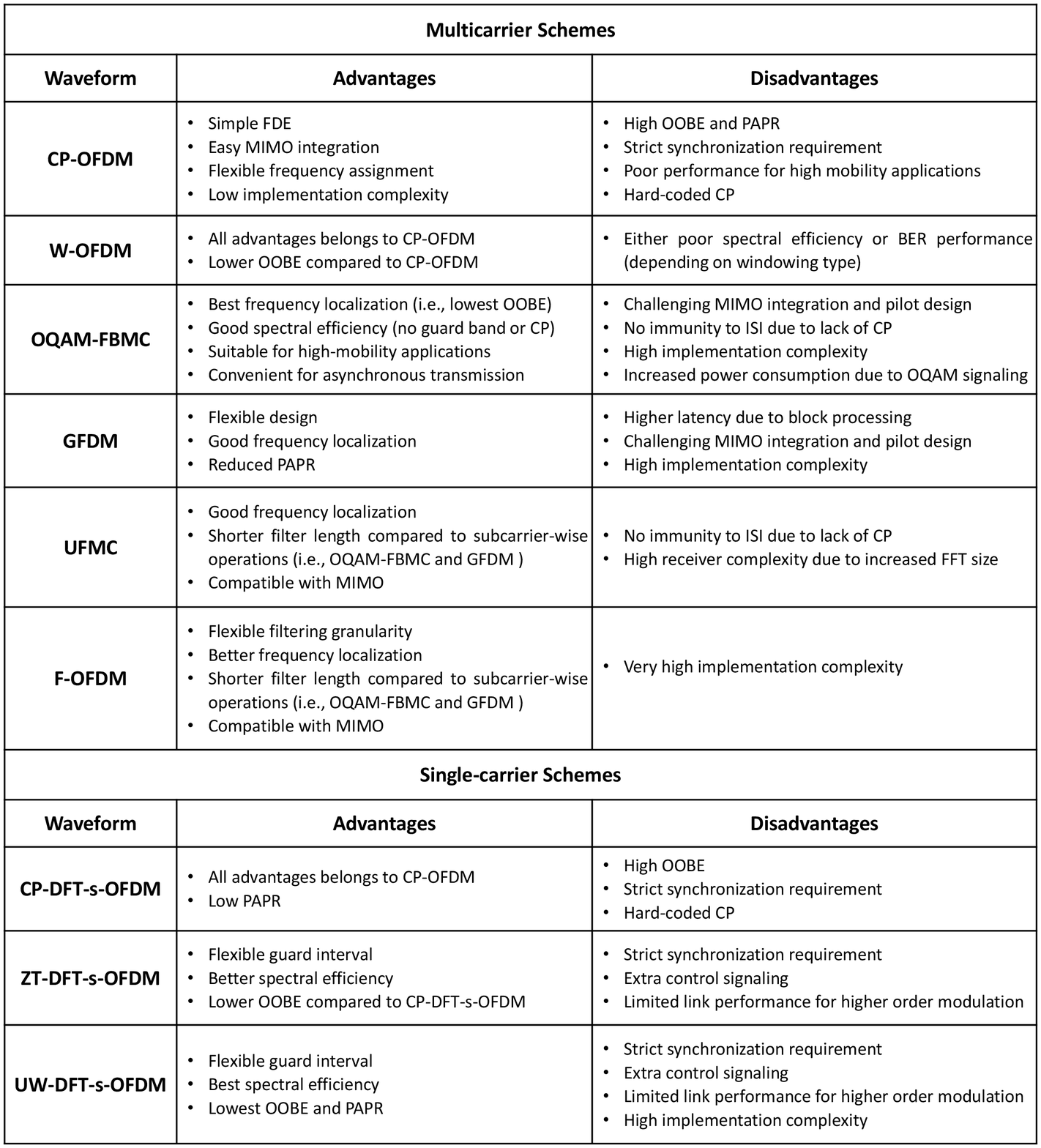}
\end{table}

\section{Conclusions \label{sec:V}}

In this chapter, the ongoing new radio design discussions are summarized,
and the major waveform candidates for 5G and beyond are presented.
The main 5G use cases are identified along with the associated critical
design requirements, and a brief description of CP-OFDM is provided
as a baseline for the new waveform discussion. The candidate waveforms
are classified considering the way the spectrum is utilized (i.e.,
single-carrier versus multicarrier), the fashion in which the signal
processing techniques are performed (i.e., windowing, subcarrier-wise
filtering, and subband-wise filtering), and the guard interval type
that is adopted (i.e., internal and external). The advantages and
disadvantages of each waveform are discussed in detail. It could be
concluded that there is no waveform that fits all requirements yet
and the physical layer should be designed considering the specific
use cases and requirements. Unlike the previous standards, the new
generations will support high flexibility to fully exploit and increase
further the potential of future communications systems.

\bibliographystyle{IEEEtran}
\bibliography{IEEEabrv,Waveform5GRef}

\vspace{4cm}

\section*{\textbf{Author Information}} 
\begin{IEEEbiography}[\includegraphics{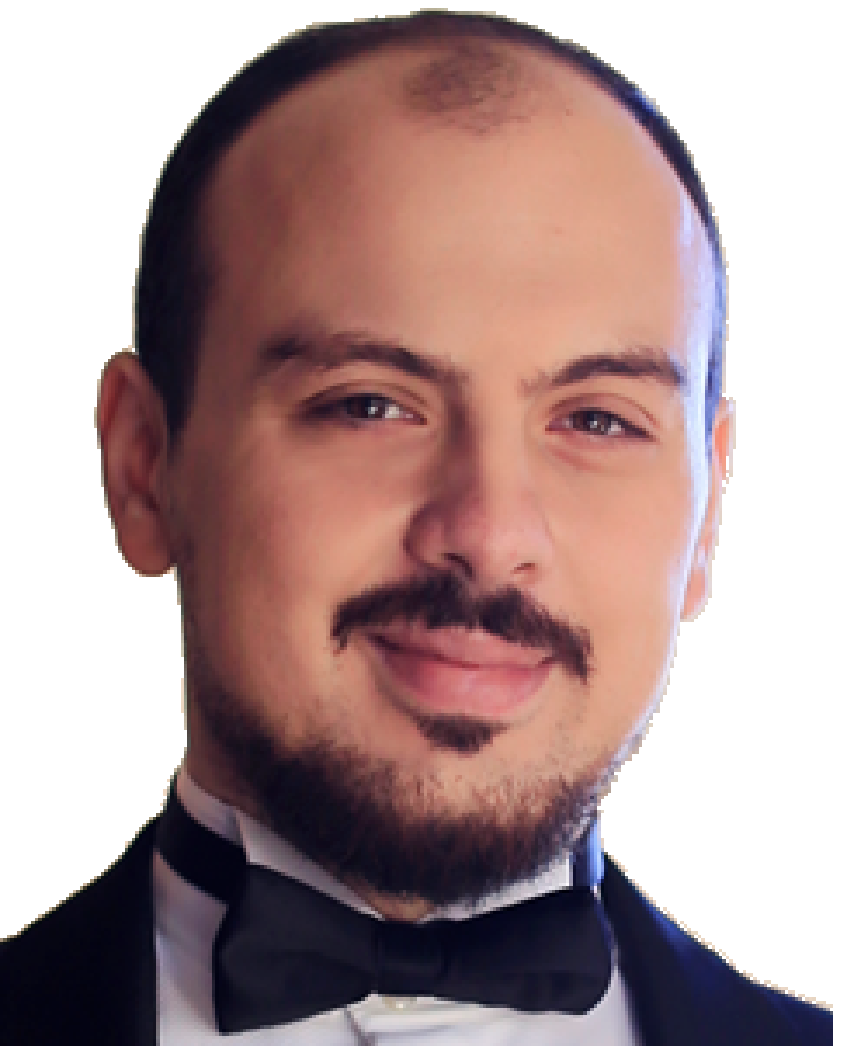}]{Ali Fatih Demir}
received the B.S. degree in electrical engineering from Yildiz Technical
University, Istanbul, Turkey, in 2011, and the M.S. degree in electrical
engineering and applied statistics from Syracuse University, Syracuse,
NY, USA in 2013. He is currently pursuing the Ph.D. degree as a member
of the Wireless Communication and Signal Processing (WCSP) Group in
the Department of Electrical Engineering, University of South Florida,
Tampa, FL, USA. His current research interests include waveform design,
multicarrier systems, \emph{in vivo} communications, and brain\textendash computer
interfaces. He is a student member of the IEEE.
\end{IEEEbiography}

\begin{IEEEbiography}[\includegraphics{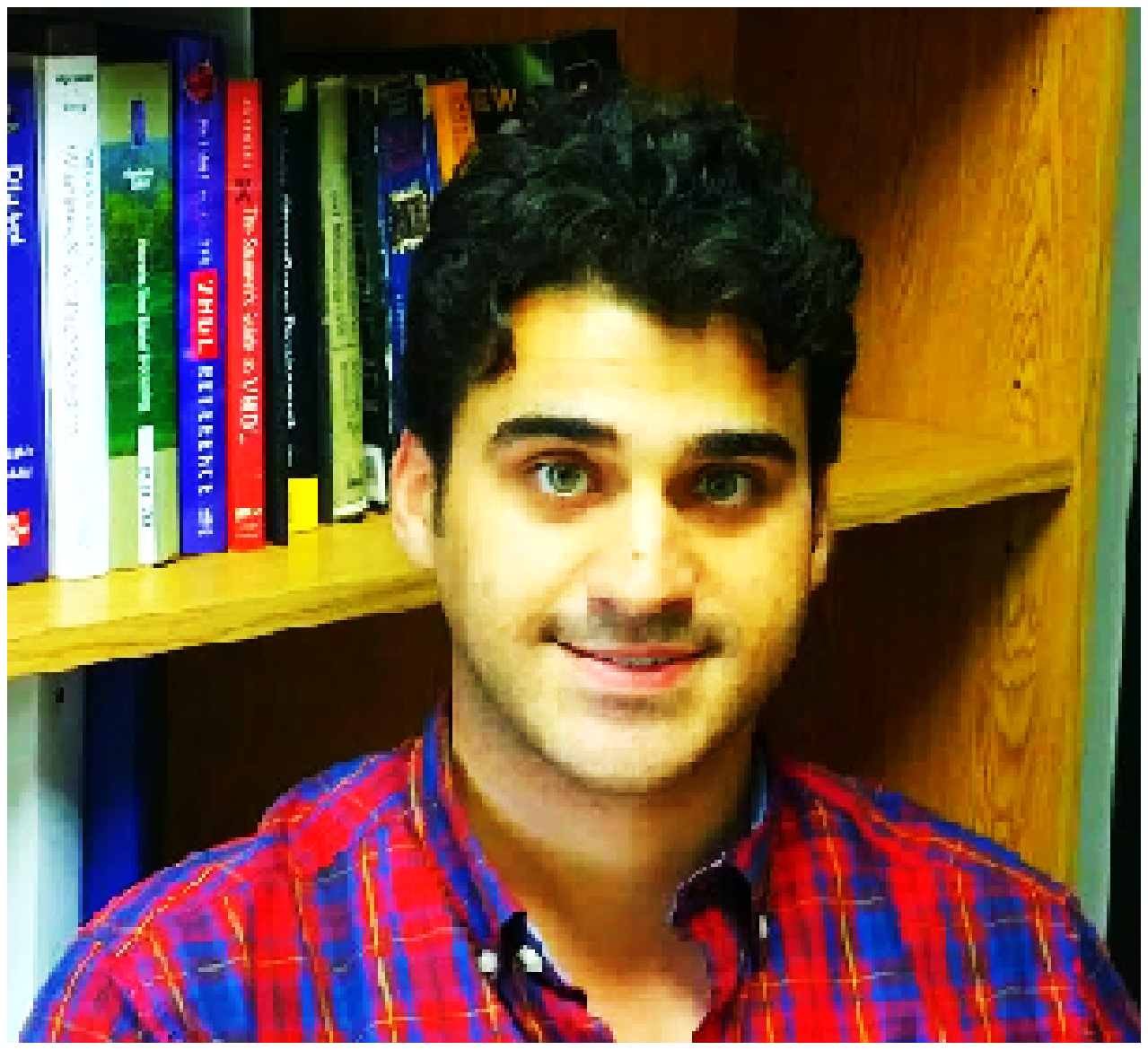}]{Mohamed H. Elkourdi}
received the B.Sc. degree in telecommunication and electronics engineering
with distinction from Applied Science University, Amman, Jordan, in
2010, and the M.S. degree in telecommunication and signal processing
from New Jersey Institute of Technology, Newark, NJ, USA in 2013.
He is currently pursuing the Ph.D. degree as a member of the Innovation
in Wireless Information Networking Laboratory (iWINLAB) in the Department
of Electrical Engineering, University of South Florida, Tampa, FL,
USA. His current research interests include waveform design, multicarrier
systems, multiple access techniques, and MIMO systems.
\end{IEEEbiography}

\begin{IEEEbiography}[\includegraphics{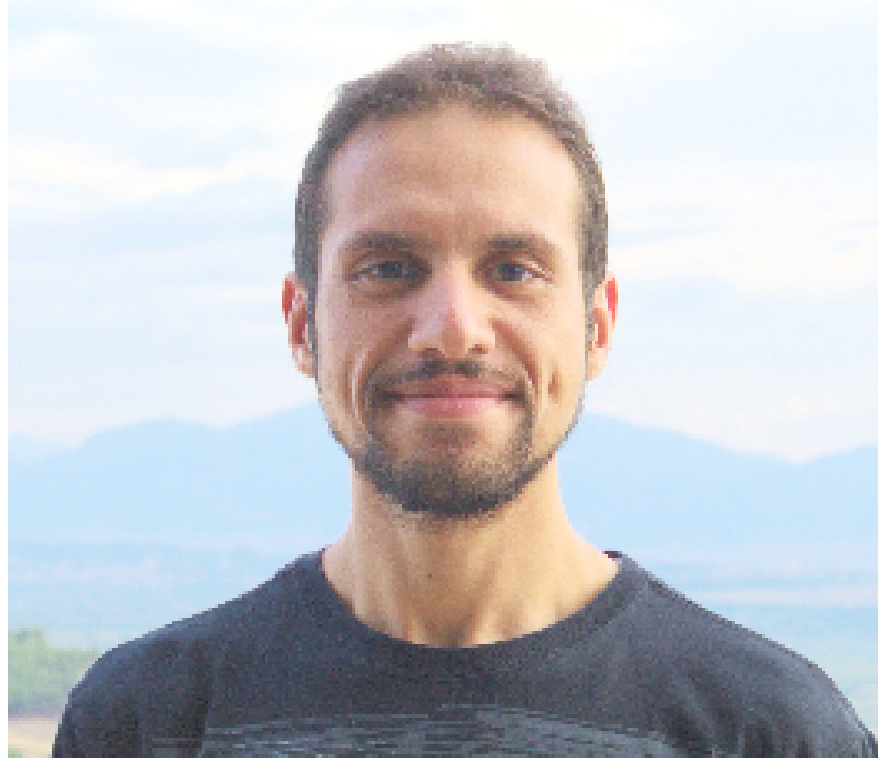}]{Mostafa Ibrahim}
received the B.Sc. degree from Ain Shams University, Cairo, Egypt
in 2010, and the M.Sc. degree from Istanbul Medipol University, Istanbul,
Turkey, in 2017, both in electronics and communication engineering.
Prior to joining the Communications, Signal Processing, and Networking
Center (CoSiNC) at Istanbul Medipol University in 2015, he was with
the Egyptian Air Force, where he was a communication engineer officer
(2011\textendash 2013), and with the Center for Nanoelectronics and
Devices (CND) at the American University in Cairo, where he worked
on on-chip energy harvesting systems optimization (2013\textendash 2014).
His current research interests include air\textendash ground channel
modeling and waveform/modulation design beyond OFDMA.
\end{IEEEbiography}

\begin{IEEEbiography}[\includegraphics{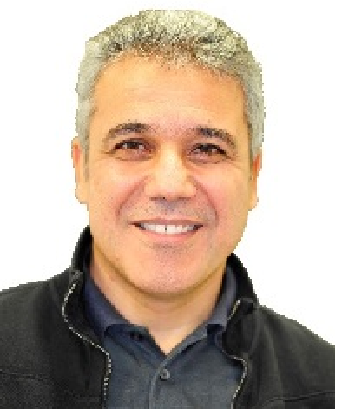}]{Huseyin Arslan}
received the B.S. degree from Middle East Technical University,
Ankara, Turkey, in 1992 and the M.S. and Ph.D. degrees from Southern
Methodist University, Dallas, TX, USA, in 1994 and 1998, respectively.
From January 1998 to August 2002, he was with the research group of
Ericsson Inc., NC, USA, where he was involved with several projects
related to 2G and 3G wireless communication systems. Since August
2002, he has been with the Department of Electrical Engineering, University
of South Florida, Tampa, FL, USA, where he is a professor. In December
2013, he joined Istanbul Medipol University, Istanbul, Turkey, where
he has worked as the Dean of the School of Engineering and Natural
Sciences. His current research interests include waveform design for
5G and beyond, physical layer security, dynamic spectrum access, cognitive
radio, coexistence issues on heterogeneous networks, aeronautical
(high-altitude platform) communications, and \emph{in vivo} channel
modeling and system design. He is currently a member of the editorial
board of the \emph{Sensors Journal} and the \emph{IEEE Surveys and
Tutorials}. He is a fellow of the IEEE.
\end{IEEEbiography}

\end{document}